\documentclass[amsmath,amssymb,aps,prl,twocolumn,nofootinbib,preprintnumbers]{revtex4}
\newcommand{\be}{\begin{equation}}
\newcommand{\bea}{\begin{eqnarray}}
\newcommand{\eea}{\end{eqnarray}}
\newcommand{\ba}{\begin{array}}
\newcommand{\ea}{\end{array}}
\newcommand{\ee}{\end{equation}}
\newcommand{\nn}{\nonumber}
\begin{document}
\title{Symmetric Orbifolds and Entanglement Entropy for Primary Excitations\\ in Two Dimensional CFT}
\author{Amir Esmaeil Mosaffa}
\affiliation{Department of Physics, Sharif University of Technology \\
P.O. Box 11365-9161, Tehran, Iran \\ 
and 
\\
Theory Group, Physics Department, CERN CH-1211 Geneva 23 SWITZERLAND\\
mosaffa@physics.sharif.edu} 

\begin{abstract} 
We use the techniques in symmetric orbifolding to calculate the Entanglement Entropy of a single interval in a two dimensional conformal field theory
on a circle which is excited to a pure highest weight state. This is achieved by  calculating the $R\acute{e}ney\ Entropy$ which is found in terms of a $2n$-point
function of primary operators, $n$ being the replica number.

\end{abstract}
\preprint{CERN-PH-TH/ 2012-219}

\maketitle

Entanglement Entropy $(EE)$ has been the subject of extensive research in the last few years. Early interests in the subject,
 \cite{'tHooft:1984re}\cite{Bombelli:1986rw}\cite{Srednicki:1993im}, came through the similarities of this quantity with the entropy of 
 black holes through the area law \cite{Bekenstein:1973ur}\cite{Hawking:1974sw}. However, $EE$ proved to be a 
powerful tool on its own for understanding
the quantum nature of physical systems 
(see for example\cite{Calabrese:2004eu, Calabrese:2005zw, Calabrese:2009qy} for reviews and references). 
Apart from being an important quantity in quantum information, $EE$ has been introduced
as a suitable order parameter for quantum phase transitions.

When there is entanglement between the degrees of freedom of two disjoint regions of space, even if they are far apart, measurements in 
one area affect those in the other instantaneously, that is, the effect is not propagated by any messenger, rather, it is the result of the
quantum structure of the system. If an observer confines himself to one of the two regions, although he has no access to 
the other, his measurements are affected by that. He makes observations in his measurements which are induced by the degrees
of freedom which are not accessible to him. $EE$ is a quantification of his lack of information about the subsystem that is accessible to him.

Lack of information can also be caused by statistical distribution of states in a system such as thermal ensembles.
In such situations $EE$ will no longer be a useful measure of quantum entanglement and thus one usually studies
this quantity when the system is in a pure state. Most of the research on this subject has focused on the case where
this pure state is the ground state of the theory. In this article we are interested in excited pure states.

After the proposal of \cite{Ryu:2006bv} (see also \cite{Ryu:2006ef,Nishioka:2009un,Takayanagi:2012kg} for reviews and references),
which gives a holographic interpretation of $EE$, there has been an even greater
interest in the subject. The present work grew out of an attempt to find the holographic description of the $EE$
of a single interval in a two dimensional CFT which has been excited by primary operators. The field theory side
of this problem has already been addressed by two different methods in \cite{Alcaraz:2011tn} and \cite{Berganza:2011mh}. 
Here we present a third method, symmetric orbifolding\cite{Lunin:2000yv},  to address this problem
which proves useful in finding its holographic description \cite{Faraji:2012}. But first some preliminaries.
\\ \\
$\it{Entanglement\ Entropy\ in\ QFT}$
\\

Suppose a physical system consists of two subsystems $A$ and $B$ and that the whole system is in a pure quantum state with
the density operator $\hat{\rho}$. Once we take the trace of the density operator over the $B$ degrees of freedom, the resultant
operator is called the $\it{reduced\ density\ operator}$ and is denoted by $\hat{\rho}_A(\equiv\ tr_B\hat{\rho})$. 
Generically $\hat{\rho}_A$ will no longer be pure and one can associate entropy to it. The $EE$ for the subsystem $A$
 is denoted by $S_A$ and is defined as the Von Neuman entropy of the reduced
density operator, $S_A\equiv-tr_A\hat{\rho}_A\ln{\hat{\rho}_A}$.

A useful mathematical quantity, called the $R\acute{e}ney\ Entropy\ (RE)$, is defined by the $replica\ trick$ as
$S_A^{(n)}\equiv \frac{1}{1-n}\ \ln{tr_A\hat{\rho}_A}^n$ such that $S_A=\lim_{n\rightarrow1}S_A^{(n)}$. Here $n$ 
is a positive integer and is called the replica number. In Quantum Field Theory (QFT)
this quantity can be represented in terms of a path integral. This is achieved by considering $n$ copies of the world volume of the original theory,
$\mathcal{M}$,
and glueing them along the entangling subspaces in a cyclic order. This results in a space, which we denote by $\mathcal{R}_n$, and which has
singularities on the boundaries of entangling subspaces. The path integral on $\mathcal{R}_n$ is denoted by $Z_{\mathcal{R}_n}$ and is defined as 
\be
Z_{\mathcal{R}_n}=\int [d\varphi(x)]\ e^{-S[\varphi]}\ ,\ \ \ x\in\mathcal{R}_n\ .\nn
\ee
This expression is proportional to $RE$. Except for some rare examples it is extremely difficult, if not impossible, to calculate $Z_{\mathcal{R}_n}$ directly.
One can go around this by transferring the geometric complexities of the world volume into the geometry of target space. That is,
one considers the original nonsingular world volume but instead introduces $n$ copies of the target space fields, 
$\varphi_i\ (i=1,2,...,n)$, on that. Instead of gluing the world volumes one now restricts the fields to satisfy certain 
conditions along the entangling subspaces
\be
Z_{res}=\int[d^n\varphi(x)]\ e^{S[\varphi_1,...,\varphi_n]}\ ,\ \ \ x\in\mathcal{M}\ , \nn
\ee
where the subscript $res$ stands for restrictions on fields. Note that these restrictions replace the nontrivial geometry of $\mathcal{R}_n$.
One way to impose the restrictions is to insert the so called $\it{twist \ operators}$
at the boundaries of entangling subspaces and calculate an unrestricted integral
\be
Z_{Twist}=\int_{unres}[d^n\varphi(x)]\ e^{S[\varphi_1,...,\varphi_n]}\ \prod \sigma_k.....\ ,\ \ \ x\in\mathcal{M}\ , \nn
\ee
where $\sigma_k$ are the twist operators that enforce restrictions through their Operator Product Expansion (OPE) with fields.
Another way of imposing the restriction is to move over to the covering space of the fields, denoted by $\mathcal{M}_C$, with a 
suitable coordinate transformation and perform the calculations on this smooth manifold
\be
Z_{\mathcal{M}_C}=\int [d\varphi(x)]\ e^{-S[\varphi]}\ ,\ \ \ x\in\mathcal{M}_C\ .\nn
\ee
On the covering space the restrictions on fields in the integration are taken care of by the geometry of $\mathcal{M}_C$.
In the following we use this last way of imposing restrictions by orbifolding techniques. These restrictions amount to identifications of target space
fields under subgroups of the symmetric group, $S_n$, and hence the name symmetric orbifolding.
This process has been worked out in full detail in \cite{Lunin:2000yv} for two dimensional theories in their ground state.
The new ingredients in our case are the primary operators which excite the theory out of its ground state.
In what follows, we focus on a generic two dimensional conformal field 
theory and consider a single entangling subspace. First a short outline of the method.
\\ \\
$\it{Symmetric\ Orbifolding}$
\\

The outline is as follows\cite{Lunin:2000yv}; suppose we start with a theory on sphere, parametrized by
$(z,\bar{z})$, with a flat metric and with two branch points of order $n$ at $u$ and $v$. Here $n$ is the 
replica number and $u$ and $v$ are the endpoints of the entangling interval.
By a coordinate transformation to $(w(z),\bar w(\bar{z}))$, which behaves as $w\approx z^{1/n}$ at branch points, one moves over to the covering sphere with the same line element (but a different metric).
By a Weyl transformation
with a conformal factor $|\frac{dw}{dz}|^{2}$,
one ends up with a third sphere with a fiducial metric $d\hat{s}^2$ which we have chosen to be flat.

As the first two spheres are related by diffeomorphism, partition functions on the two are equal. This in turn is related to the partition function on the third 
sphere by the exponent of the $\it{Liouville}$ action imposed by Weyl anomaly.
A careful calculation of this action results in the known expression for RE.
\\ \\
$\it{The\ case\ for\ excitations}$
\\

We now wish to calculate the $RE$ for a CFT on a sphere with a single branch cut. We further assume that the state we start with is a
highest weight state with weights $(h,\bar{h})$. To proceed we use the following parameterisation for the sphere
\bea
ds^2&=&dz\ d\bar{z}\ ,\ \ \ \ z<\frac{1}{\delta}\ ,\\
&=&d\tilde{z}\ d\bar{\tilde{z}}\ , \ \ \ \ \tilde{z}<\frac{1}{\delta}\ ,\nn\\
\tilde{z}&=&\frac{1}{\delta^2}\frac{1}{z}\nn\ .
\eea
We call this the $z$-sphere for which we have chosen a flat metric with a regularisation parameter $\delta$. 
Without loss of generality we choose the branch points as following
\be
u=ae^{\frac{i}{2}(\pi+\theta)}\ \ \ ,\ \ \ v=ae^{\frac{i}{2}(\pi-\theta)}\ .
\ee
In order for the theory to be in a highest weight excitation, we create  the asymptotic $\it{in}$ and $\it{out}$ states by putting 
the corresponding operator $\mathcal{O}(z,\bar{z})$ at $z=\bar{z}=0$ and $\tilde{\mathcal{O}}(\tilde{z},\bar{\tilde{z}})$
at $\tilde{z}=\bar{\tilde{z}}=0$ where
\be\label{Conj}
\tilde{\mathcal{O}}(\tilde{z},\bar{\tilde{z}})=\mathcal{O}(z,\bar{z})z^{2h}\bar{z}^{2\bar{h}}\ \delta^{2(h+\bar{h})}\ ,
\ee
and $\mathcal{O}$ is a primary operator with weights $(h,\bar{h})$. Note that the operator $\tilde{\mathcal{O}}$ is defined on the north pole cap
which is parameterised by $\tilde{z}$ and $\bar{\tilde{z}}$. One can equivalently create the $\it{out}$ state at the north pole by introducing the adjoint operator as
\footnote{In the following, as a shorthand, we will be rather sloppy with this notation and denote the $\it{out}$ state by $\tilde{\mathcal{O}}$ in the $z$ and $\bar{z}$ argument.
So, for example, an $\it{out}$ state in the north pole will be denoted by $\tilde{\mathcal{O}}(\infty)$.}
\bea
\lim_{z,\bar{z}\rightarrow0}\mathcal{O}^{\dag}(\bar{z},z)&=&\lim_{\tilde{z},\bar{\tilde{z}}\rightarrow0}\tilde{\mathcal{O}}(\tilde{z},\bar{\tilde{z}})\nn\\
&=&\lim_{z,\bar{z}\rightarrow\infty}\mathcal{O}(z,\bar{z})z^{2h}\bar{z}^{2\bar{h}}\ \delta^{2(h+\bar{h})}\ .
\eea

We should now make an $n$-sheeted Riemann sphere by appropriately gluing $n$ copies of the above spheres along the branch cuts and calculate the path integral on 
the resulting manifold. As stated before, one can equivalently calculate a restricted path integral of $n$ copies of the theory on a single sphere. We denote such a quantity by 
$tr\rho_{\mathcal{O}}^n(\theta)$ which is defined as
\bea\label{RhoO}
tr\rho_{\mathcal{O}}^n(\theta)&\equiv&\frac{\int_{res} {[d^n\varphi]}e^{-S[\varphi_1,...,\varphi_n]}\prod_{i=1}^{n}\mathcal{O}_i(0)\tilde{\mathcal{O}}_i(\infty)}
{\big[\int [d\varphi]e^{-S[\varphi]}\mathcal{O}(0)\tilde{\mathcal{O}}(\infty)\big]^n}\ ,\nn\\
\eea
where $\theta$ determines the entangling interval and the expression in the denominator is a normalization factor which ensures $tr\rho_{\mathcal{O}}=1$.

As explained in \cite{Lunin:2000yv}, the above path integral can be properly defined by cutting out circular holes of radius, say, $\epsilon$ around the branch points
as well as the infinity. One should then specify the proper boundary conditions for fields
along the edges of the holes. The precise way of performing this procedure has been carried out in \cite{Lunin:2000yv} with full details which we mostly skip. 
There are two new ingredients
in our case which should be addressed. The first one is that unlike the case of \cite{Lunin:2000yv} where the unity operator is inserted in $z=\infty$, we insert the primary operator
$\tilde{\mathcal{O}}$. Secondly, we are also inserting $\mathcal{O}$ at $z=0$. We address these issues in the following.

Let us define the covering space, the $w$-sphere, through the map
\be\label{Diff}
\frac{z-u}{z-v}=\frac{1}{1-(\frac{w-1}{w+1})^n}\ .
\ee
Near the branch points, the map behaves as $z\approx t^n$.
On this sphere the metric is induced through the map as 
\be
ds^2=\frac{dz}{dw}\frac{d\bar{z}}{d\bar{w}}dwd\bar{w}\ ,
\ee
and the regularisation parameter is also found as $\delta'=a\delta\sin{(\theta/2)}/n$. There are several holes on this sphere.
Two of these are the images of the branch points. There are also $n$ holes coming from the images of $z=\infty$ and  $n$ holes from $z=0$.

The prescription of filling these holes should be such that upon path integration inside the holes we should end up with the desired states at the edges.
As for the branch points, one such prescription is given in \cite{Lunin:2000yv}. This is roughly filling the holes with disks of a flat metric which is continuously matched with the metric outside the hole over the edge. There is however a curvature concentration along the edge which is expected because around each branch point there is a cone with an excess angle.

The holes coming from $z=\infty$ are naturally filled in by the images of $|\tilde{z}|<1/\delta$ and those coming from $z=0$ by the images of $|z|<\epsilon$. The vertex operators $\tilde{\mathcal{O}}$ and $\mathcal{O}$ guarantee the desired wave functionals at the edges.

We now have a closed surface, the $w$-sphere, on which we want to calculate a certain path integral in presence of operator insertions.
As usual, we choose a fiducial metric, which for simplicity, we take it 
to be a flat metric with an arbitrary regularisation parameter\footnote{We take $\hat{\delta}<\delta'$ as a convenient choice for reasons to follow.}, $\hat{\delta}$.
This is achieved by performing a Weyl transformation on the $w$-sphere, with a factor $|dz/dw|^2$, to make the metric flat. We then introduce the coordinate $t=w$ but
choose $\hat{\delta}$  as the regularisation parameter which defines $\tilde{t}=1/(\hat{\delta}^2t)$.

The $t$-sphere thus defined is where we perform our calculations. Note that the path integrals on the different spheres are schematically related as
\bea
\int_{res}[d^n\varphi(z,\bar{z})]e^{-S[\varphi_1,...,\varphi_n]}...=\int[d\varphi(w,\bar{w})]e^{-S[\varphi]}...\nn\\
=e^{S_L}\int[d\varphi(t,\bar{t})]e^{-S[\varphi]}...\ ,
\eea
where dots stand for possible insertions, in case of which, the appropriate transformation factors should be included.
$S_L$ in the above relation is the $\it{Liouville}$ action coming from Weyl anomaly and is defined as
\be\label{Liou}
S_L=\frac{c}{96\pi}\int dt^2\sqrt{g}[ \partial_{\mu}\phi\partial_{\nu}\phi g^{\mu\nu}+2R\phi]\ ,
\ee
where $c$ is the central charge of the theory and 
\be
e^{\phi}=|\frac{dz}{dt}|^2\ .
\ee
The transformation(\ref{Diff}) turns the restricted path integral into an unrestricted one. This is so because 
the $w$-sphere, being an $n$-fold cover of the $z$-sphere, automatically enforces the desired identifications of fields along the branch cut.

As for the insertions, we note that
the sequence of  the $\it{Diff \times\ Weyl }$ transformations, keeping the metric invariant, is a conformal transformation on the whole. 
Therefore for the primary fields $\mathcal{O}$, we have
\be\label{Tran}
\mathcal{O}(z,\bar{z})=\bigg(\frac{dt}{dz}\bigg)^h\bigg(\frac{d\bar{t}}{d\bar{z}}\bigg)^{\bar{h}}\ \mathcal{O}(t,\bar{t})\ ,
\ee
and a similar one for $\tilde{\mathcal{O}}$. 

Putting everything together we find that 
\be\label{Summ}
tr\rho_{\mathcal{O}}^n(\theta)=e^{S_L}\frac{Z_t}{Z_z^n}\ \mathcal{T}\ 
\frac{\langle \prod_{k=0}^{n-1}\mathcal{O}(t_k)\tilde{\mathcal{O}}(t'_k)\rangle_t}    {\langle\mathcal{O}(0)\tilde{\mathcal{O}}(\infty)\rangle_z^n}\ ,
\ee
where $\mathcal{T}$ is the transformation factor for operators, a product of those appearing in ($\ref{Tran}$), and $t_k$ and $t'_k$ 
are the images of $z=0$ and $\tilde{z}=0$ respectively. $Z$ stands for partition function and the subscripts $z$ and $t$ denote on which 
sphere the corresponding quantities are calculated.

We now have all the ingredients to perform our calculations. First start with the Liouville part. 
It turns out that there are three different contributions to $S_L$ (see \cite{Lunin:2000yv} for details).
One is coming from the kinetic term in the  $1/{\delta'}<|t|<1/\hat{\delta}$ region and the other comes 
from the curvature ring at $|t|=1/\hat{\delta}$ $\footnote{In \cite{Lunin:2000yv} these are denoted by $S_L^{(2)}$ and $S_L^{(3)}$ respectively.}$. These two  sum up to an expression which only depends on regulators $\delta$ and $\hat{\delta}$ and which cancels out in the end.

The main contribution to $S_L$, which we call $S_L^{(1)}$, comes from the region which is bounded between the edges of the holes, on the one hand,  and $|t|=1/\delta'$ on the other.
Given that the metric is flat in this region, the only contribution to $(\ref{Liou})$ comes from the Kinetic term which we can turn into a boundary term
\be\label{SL1}
S_L^{(1)}=\frac{c}{96\pi}[i\int dt \ \phi\partial_t\phi+c.c.]\ ,
\ee
where the integration is along the boundaries of the region. The calculation of this integral for the branch points as well as the images of $z=\infty$ is 
identical to those in \cite{Lunin:2000yv} and nothing changes. As for the images of $z=0$, it is straightforward to show that near such points
\be
\phi=\log{|\frac{dz}{dt}|^2}\approx \log{(c+a(t-t_0))}\ ,\   \partial_t \phi\approx \frac{1}{c+a(t-t_0)}\ ,\nn
\ee
where $a$ and $c$ are constants and $t_0$ is any of the images of $z=0$. The integral $(\ref{SL1})$ for these values in the limit $t\rightarrow t_0$ will obviously be zero
and there is no contribution to $S_L$ from the images of $z=0$.
The upshot is that as far as $S_L$ is concerned, operator insertions have no effect. Recalling that $tr\rho^n=e^{S_L}Z_t/Z_z^n$, and consulting ($\ref{Summ}$),
this statement leads to 
\be
\frac{tr\rho_{\mathcal{O}}^n(\theta)}{tr\rho^n(\theta)}= \mathcal{T}\ \frac{\langle \prod_{k=0}^{n-1}\mathcal{O}(t_k)\tilde{\mathcal{O}}(t'_k)\rangle_t} 
 {\langle\mathcal{O}(0)\tilde{\mathcal{O}}(\infty)\rangle_z^n}\equiv \mathcal{F}^{(n)}_{\mathcal{O}}(\theta)\ .
\ee

We now find the effect of vertex operators, i.e., calculate the factor $\mathcal{T}$ appearing in ($\ref{Summ}$).
On $t$-sphere the branch points $z=(u,v)$ are mapped to $t=(-1,1)$.
The point $z=0$, on the other hand, is mapped to $n$ points which we denote by $t_k$
\be
t_k=-i\cot{\bigg(\frac{\theta+2\pi k}{2n}\bigg)}\ , \ \ \ k=0,1,...,n-1.
\ee
The images of $z=\infty$ are denoted by $t'_k$ where
\be
t'_k=-i\cot{\big(\frac{\pi k}{n}\big)}\ , \ \ \ k=0,1,...,n-1.
\ee
Note that $t'_0=\infty$.

To find $\mathcal{T}$, we need to calculated $dt/dz$ at $t=t_k\  (k=0,1,...,n-1)$, $dt/d\tilde{z}$ at $t=t'_k\  (k=1,...,n-1)$ and finally $d\tilde{t}/d\tilde{z}$ at $t=t'_0$.
One can then write $\mathcal{T}$ as a product of
\be
\mathcal{T}=\bigg(\frac{d\tilde{z}}{d\tilde{t}}\bigg|_{t'_0}\times\prod_{k=1}^{n-1}\frac{d\tilde{z}}{dt}\bigg|_{t'_k}\times\prod_{k=0}^{n-1}\frac{dz}{dt}\bigg|_{t_k}\bigg)^{-h}\ ,
\ee
times a similar expression but with $(z,t,h)$ replaced with $(\bar{z},\bar{t},\bar{h})$. One finds that
\bea
\frac{dz}{dt}\bigg|_{t_k}&=&\frac{an}{\sin(\theta/2)}\sin^2\bigg(\frac{\theta+2\pi k}{2n}\bigg)\ ,\  k=0,1,...,n-1\ ,\nn\\ 
\frac{d\tilde{z}}{dt}\bigg|_{t'_k}&=&\frac{n}{\delta^2a\sin(\theta/2)}\sin^2\big(\frac{\pi k}{n}\big)\ , \ \ \ k=1,2,...,n-1\ ,\nn\\
\frac{d\tilde{z}}{d\tilde{t}}\bigg|_{t'_0}&=&\big(\frac{\hat{\delta}}{\delta}\big)^2\frac{n}{a\sin(\theta/2)}\ ,
\eea
which gives the final result as
\bea\label{Main}
\mathcal{F}^{(n)}_{\mathcal{O}}(\theta)&=&\bigg(\delta^n\ \frac{ 2^{(n-1)}}{\hat{\delta}}\ \frac{\sin^{(n-1)}(\theta/2)}{n^{(n+1)}}\bigg)^{2(h+\bar{h})}\nn\\
&\times&
\frac{\langle \prod_{k=0}^{n-1}\mathcal{O}(t_k)\tilde{\mathcal{O}}(t'_k)\rangle_t} 
{\langle\mathcal{O}(0)\tilde{\mathcal{O}}(\infty)\rangle_z^n}\ .
\eea
This is our main result. Let us find an approximation to this formula in the limit $\theta\ll2\pi$. Recall that
\be
\mathcal{O}(t,\bar{t})\tilde{\mathcal{O}}(0,0)=\frac{1}{t^{2h}\bar{t}^{2\bar{h}}}
[1+\mathcal{Q}_{\Delta,\bar{\Delta}}t^{\Delta}\bar{t}^{\bar{\Delta}}+\dots]\ ,
\ee
where dots in the second line stand for higher powers of $t$ and $\bar{t}$ and $\mathcal{Q}_{\Delta,\bar{\Delta}}$
is the operator with smallest dimensions, $(\Delta,\bar{\Delta})$, in the OPE. This gives
\bea
\prod_{k=0}^{n-1}\mathcal{O}(t_k)\tilde{\mathcal{O}}(t'_k)&=&\bigg(\frac{\hat{\delta}}{2^{(n-1)}}\ \frac{n\sin{(\theta/2)}}{\sin^n(\theta/2n)}\bigg)^{2(h+\bar{h})}\nn\\
&\times&\big(1+O(\theta^{(\Delta+\bar{\Delta})})\big)\ .
\eea
We also recall that
\be
\langle\mathcal{O}(0)\tilde{\mathcal{O}}(\infty)\rangle_z=\delta^{2(h+\bar{h})}\ .
\ee
Putting everything together
\be
\mathcal{F}^{(n)}_{\mathcal{O}}(\theta)=1+\frac{h+\bar{h}}{3}\big(\frac{1}{n}-n\big)(\frac{\theta}{2})^2+O(\theta^{(\Delta+\bar{\Delta})})\ ,
\ee
which is in complete agreement with the result in \cite{Alcaraz:2011tn} in the same limit.

To further compare our result to those in \cite{Alcaraz:2011tn} and \cite{Berganza:2011mh}, we define yet another coordinate $s$ by
\be
\frac{t+1}{t-1}=e^{is}\ ,
\ee
which maps the $t$-sphere into a cylinder with unit radius. Upon this transformation, the expression in ($\ref{Main}$) will find the
following form
\bea
\mathcal{F}^{(n)}_{\mathcal{O}}(\theta)&=&\frac{\langle \prod_{k=0}^{n-1}\mathcal{O}(\frac{\theta+2\pi k}{n})\tilde{\mathcal{O}}(\frac{2\pi k}{n})\rangle_{cy}} 
{\langle\mathcal{O}(\frac{\theta}{n})\tilde{\mathcal{O}}(0)\rangle_{cy}^n}\nn\\
&=&\frac{n^{-2n(h+\bar{h})}\langle \prod_{k=0}^{n-1}\mathcal{O}(\frac{\theta+2\pi k}{n})\tilde{\mathcal{O}}(\frac{2\pi k}{n})\rangle_{cy}} 
{\langle\mathcal{O}(\theta)\tilde{\mathcal{O}}(0)\rangle_{cy}^n}\ ,\nn\\
&&
\eea
where in the second line we have scaled the arguments in the denominator by a factor $n$.
This is the result found in \cite{Alcaraz:2011tn} and \cite{Berganza:2011mh}.

$EE$ is now immediately obtained by calculating $S=\partial/\partial_n\ tr\rho_{\mathcal{O}}^n(\theta)$ at $n=1$. Noting that 
$tr\rho_{\mathcal{O}}(\theta)=\mathcal{F}^{(1)}_{\mathcal{O}}(\theta)=1$, we find
\be
S_{\mathcal{O}}(\theta)=S_{GS}(\theta)-\frac{\partial}{\partial_n}\mathcal{F}^{(n)}_{\mathcal{O}}(\theta)|_{n=1}\ .
\ee
Given the Gauge/Gravity duality, \cite{Maldacena:1997re}\cite{Gubser:1998bc}\cite{Witten:1998qj}, it is a fair question to ask for a holographic analogue of the above calculations and results. 
This will amount to identifying the gravitational counterparts of the procedure used here. 

In the original proposal of \cite{Ryu:2006bv}, the
distinct role of the branch points on the boundary result in a desirable geometric realisation of $EE$ in terms of minimal lengths.
This is intuitively understood by extending the curvature concentration at the branch points into the bulk.
Using the covering space, as is the case in this letter, this geometric intuition is lost because the singularities
at branch points are smoothed out. One should then find the bulk geometry which corresponds to the
covering space in field theory. Exciting the theory into primary states will then be a matter of turning on appropriate bulk fields.
This is the subject of \cite{Faraji:2012}.
\\

I would like to thank Ali Davody for collaboration on the early stages of this work.
I would also like to thank Amin Faraji for discussions and collaboration on related projects and
Mohsen Alishahiha for discussions and a reading of the draft.

\end{document}